\documentclass[aps,prb,showpacs,twocolumn]{revtex4}
\usepackage{graphicx}
\usepackage{epsfig}
\usepackage{amsmath}
\usepackage{graphics}
\usepackage{epsfig}
\begin{document}
\title{Nonequilibrium Fock space for the electron transport problem}

\author{D. S. Kosov }
\address{
Department of Physics and  Center for Nonlinear Phenomena and Complex Systems,
Universit\'e Libre de Bruxelles, 
Campus Plaine, CP 231, 
Blvd du Triomphe, 
B-1050 Brussels, 
Belgium   }

\email{dkosov@ulb.ac.be}
 

\begin{abstract} 
Based on the formalism of thermo field dynamics we propose a concept of nonequilibrium Fock space and nonequilibrium quasiparticles for quantum many-body system in nonequilibrium steady state.
We develop a general theory as well as demonstrate the utility of the approach on the example of electron transport through the interacting region.
The proposed approach is compatible with advanced quantum  chemical methods  of electronic structure calculations such as coupled cluster theory and configuration interaction.
\end{abstract}

\maketitle

Accurate and reproducible measurements of current-voltage characteristics
have been reported for single molecules.\cite{venkataraman:458,ruitenbeek08} The experiments have revealed a wealth of new
interesting physical phenomena in molecules such as Coulomb blockade, Kondo effects,
negative differential resistance, vibronic effects and local heating, as well as switching
and hysteresis.\cite{natelson04,Elbing05} 
Much of the theoretical and computational studies  have been  based on Keldysh  nonequilibrium GreenÕs functions,\cite{Taylor01,arnold:174101,galperin:035301,dahnovsky:014104,li:035415}
but the understanding of fundamental mechanisms of quantum transport 
in nanoscale molecular systems also
requires the development of new theoretical approaches to nonequilibrium interacting many-particle
quantum systems.
In this paper,  we  develop the transport theory,
which is based on  some more recent advances in quantum field theory of nonequilibrium states.

To carry out our program we will use the formalism of thermo field dynamics (TFD).\cite{umezawa} The TFD is one of  the approaches to quantum field theory at finite temperature.  It is an operator based formalism which is complimentary to Matsubara and nonequilibrium  Green's functions. 
TFD has been successfully applied to nuclear physics,\cite{Hatsuda89,avdeenkov96,dzhioev04} particle physics,\cite{Henning1995235} and problems of  statistical mechanics,\cite{suzuki86} although its applications to chemical physics  have been very limited.  As  we  discuss in this paper, TFD can also be naturally used to describe nonequilibrium  steady states of many-particle quantum systems, in general, and quantum transport of electrons through the interacting region, in particular.

To begin, we first  summarize relevant for us aspects of TFD. A nonequilibrium quantum system can be described in terms of density matrix $\rho(t)$ that is obtained as a solution of Liouville equation
($\hbar=1$):
\begin{equation}
i \dot{\rho}(t) = \left[H, \rho(t) \right].
\label{gamma}
\end{equation}
We can introduce the square root of the density matrix 
($\rho(t) = \gamma(t) \gamma^{\dagger}(t)$), which satisfies the same time evolution equation as the density matrix:
\begin{equation}
i \dot{\gamma}(t) = \left[H, \gamma(t) \right].
\label{gamma2}
\end{equation}
The average value of an operator at particular moment $t$ can be computed as
\begin{equation}
\langle A(t) \rangle  =   \mbox{Tr} \left( \rho(t) A \right) =\mbox{Tr} \left( \gamma^{\dagger}(t) A \gamma(t) \right).
\label{neav}
\end{equation}
The TFD defines a  time-dependent wavefunction $|\psi(t)\rangle $
in such a way that the matrix element over this wavefuncion $\langle \psi(t) | A| \psi(t) \rangle $
agrees with nonequilibrium average (\ref{neav}). We assume here for simplicity that the eigenvalues of the system Hamiltonian $H$ are discrete: 
$
H | n \rangle = E_n | n \rangle , \; \langle m | n \rangle = \delta_{nm}
$.
TFD introduces a fictitious system (denoted by tilde) which is identical copy of the original system:
$
\tilde{H} | \tilde{n} \rangle = E_n | \tilde{n} \rangle , \;  \langle \tilde{m} | \tilde{n} \rangle = \delta_{nm}
$.
Then we can write wavefunction for nonequilibrium state as
$
|\psi(t)\rangle
=\sum_{nm} \gamma_{mn}(t)  |n\rangle \otimes  |\tilde{m}\rangle,
$
where $ \gamma_{nm} = \langle n |\gamma | m \rangle $  are matrix elements of the square root of the density matrix
computed in the basis of eigenvectors of the Hamiltonian of the system.
The average of operators defined in our physical space  $A= A\otimes \tilde{I}$
over this wavefunction gives exactly nonequilibrium average (\ref{neav}).   The Liouville equation (\ref{gamma2})
is equivalent to the time-dependent Schr\"odinger equation
\begin{equation}
i  |\dot{\psi}(t)\rangle = {\cal H} |\psi(t)\rangle,
\label{nesh}
\end{equation}
where  ${\cal H} = H- \tilde{H} $ is the so-called thermal Hamiltonian of the system.\cite{umezawa}
TFD should not be confused with simple notational reformulation of the Liouville equation. There are profound physical 
differences between the two approaches, which will be discussed in the end of the next paragraph.

Let us consider the quantum system which is brought to nonequilibrium by  interaction with one or  several macroscopic thermal baths or particle reservoirs.  
The total Hamiltonian  $ H=H_S+H_B +H_{SB} $ includes the system Hamiltonian $H_S$,  the  Hamiltonian of the baths $H_B $, and the system-bath interaction $H_{SB}$ and it is assumed to be time-independent.
The corresponding thermal Hamiltonian is
\begin{equation}
 {\cal H} = H - \tilde{H}= {\cal H}_S + {\cal H}_B +  {\cal H}_{SB}.
 \label{th0}
\end{equation}
Here we denote all thermal baths collectively by $B$ and all operators which are defined in the bath Hilbert space will be also denoted  by $B$, for example, $A_B$.
The equation (\ref{nesh}) should be solved 
subject to the special boundary conditions for operators acting in  Hilbert subspace of the baths. Namely,
at $t \rightarrow -\infty$, when 
 the coupling between baths and the system $H_{SB} =0$, the baths were prepared in the canonical ensembles at temperatures $T_B$:
 \begin{equation}
\lim_{t\rightarrow -\infty}\langle \psi(t) | A_B |\psi (t) \rangle = \frac{1}{Z_B} \sum_{b \in B} e^{-E_b/T_B} \langle b | A_B | b \rangle,
\label{ensemble-av}
\end{equation}
where 
$Z_B $ is the partition function of bath $B$ and $ | b \rangle $ are  eigenstates with eigenergies $E_b$ of the Hamiltonian of the bath $B$.
Let us consider the particular  case, which will be very important for our development of transport theory, namely, we consider the baths containing 
non-interacting fermions
$
H_B= \sum_B \sum_{b \in B} \varepsilon_b a_b^{\dagger} a_b$. Here $a^{\dagger}_b$($a_b$) creates (destroys) an electron in the state $b$ of the bath $B$.
 As was demonstrated by Umezawa\cite{umezawa}, in this case 
$|\psi(t\rightarrow -\infty) \rangle $ becomes a vacuum for operators $\beta_b, \tilde{\beta}_b$,
 which are related to the original bath particle creation and annihilation operators via thermal canonical Bogoliubov transformation:
  \begin{eqnarray}
\left. \begin{array}{c}
  \beta_b  
= \sqrt{1-n_b}\; a_b -  \sqrt{n_b }\; \tilde{a_b}^{+}  \\
\tilde{\beta}_b= \sqrt{1-n_b} \; \tilde{a_b}+  \sqrt{n_b }\;  a_b^{+}
\end{array}
\right\}\;, b\in B
\label{bogol}
\end{eqnarray}
Here $n_b =  (1+e^{ (\varepsilon_b -\mu_B)/T_B})^{-1}$ is the Fermi-Dirac occupation number for level $b$ in the bath and 
the canonical Bogoliubov transformation should be performed separately for all baths $B$.
Now we are ready to discuss  the important advantages of the TFD approach over the Liouville equation. 
TFD operates in the space of the distribution amplitudes (Hilbert space),  whereas the Liouville operator is defined in the space of distribution functions. As a consequence,  unitary representations of states and the definition of quasiparticles as unitary transformation of the original particles are not possible in the Liouville space. The  TFD Hilbert space is also specially entangled and compressed in order to include 
asymptotic nonequilibrium boundary conditions, in other words,  all equilibrium or nonequilibrium statistical mechanics are embedded directly in the Hilbert space of the system in the TFD.

Let us now consider a system in  nonequilibrium steady state.
 A steady state is established by the delicate balance between irreversible processes and the driving forces produced by the reservoirs.
 Nonequilibrium steady-state systems are ubiquitous in nature and their
theoretical description  has been a challenging fundamental problem for many years.\cite{gaspard06}
 We define a nonequilibrium steady state 
as an asymptotic  state ($t\rightarrow 0$) of a finite quantum system which was placed 
at $t=- \infty$ into contact with several different macroscopic  thermal baths or  particle reservoirs.\cite{gelin:022101}
We write the  total thermal Hamiltonian in the following form
\begin{equation}
 {\cal H}(t,\eta) = H - \tilde{H}= {\cal H}_S + {\cal H}_B + \exp(\eta t) {\cal H}_{SB},
\end{equation}
where $\eta$ is a small positive quantity. The  system is isolated from the baths  at infinite past, so the system  and the baths are
initially prepared in thermal equilibrium: 
\begin{equation}
| \psi (-\infty)\rangle = | \psi_S\rangle \otimes | \psi_B\rangle,
\label{init}
\end{equation}
where 
\begin{equation}
| \psi_S\rangle = \frac{1}{\sqrt{Z_S}} \sum_s \exp(-E_s/2T_s) |s \rangle \otimes |\tilde{s} \rangle,
\end{equation}
\begin{equation}
| \psi_B\rangle = \prod_B \frac{1}{\sqrt{Z_B}} \sum_{b \in B} \exp(-E_b/2T_B) |b \rangle \otimes |\tilde{b} \rangle,
\end{equation}
and obviously 
$  ({\cal H}_S  + {\cal H}_B )| \psi (-\infty)\rangle  = 0$.
 What nonequilibrium  state $ |\psi \rangle $ 
will develop adiabatically from $ | \psi(-\infty) \rangle$  (\ref{init}) if we turn on the system bath coupling? If we follow Gell-Mann and Low derivation \cite{gellmann-low}
and  perform the calculations with finite adiabaticity parameter $\eta$, it is possible to show that
\begin{equation}
( {\cal H}_S + {\cal H}_B + {\cal H}_{SB} ) |{\psi}_{\eta} \rangle =  i \eta   g \frac{\partial}{\partial g} | {\psi}_{\eta} \rangle,
\label{gl}
\end{equation}
where ${\cal H}_{SB} $ is proportional to a coupling strength $g$ and $|{\psi}_{\eta} \rangle = \hat{T} \exp(-i \int_{-\infty}^0 d\tau {\cal H}(\tau,\eta))| \psi (-\infty)\rangle $.
Contrary to the ordinary quantum mechanics nonequilibrium wavefunction does not acquire an infinite phase. Therefore, if  the system-bath coupling is switched on adiabatically (i.e. $\eta \rightarrow 0$), then the r.h.s of eq.(\ref{gl})
becomes zero.  It is mathematically and physically similar to the observation that for the classical nonequilibrium systems the slowly switching the interaction on changes  the distribution but not the eigenvalue of the Liouville operator.\cite{aronson66} Therefore, if we switch on the coupling between system and baths adiabatically slow, the initial equilibrium distribution (\ref{init}) will evolve to  the nonequilibrium steady state wavefunction  $|\psi\rangle $ which is an eigenstate of the total thermal Hamiltonian with zero eigenvalue.
Let us summarize the formal results: (i) the TFD time-dependent Schr\"odinger equation (\ref{nesh}) is completely equivalent to the Liouville equation for nonequilibrium density matrix, (ii) $|\psi(t)\rangle$ obtained from   (\ref{nesh}) contains all information about the system necessary to compute any nonequilibrium averages, (iii) for the system connected to the baths of non-interacting particles, the  boundary condition that the system-environment is decoupled in the initial state are equivalent to the requirement that $ |\psi(t) \rangle $ becomes a vacuum for annihilation operators defined via thermal Bogoliubov transformation for the bath particles (\ref{bogol}), (iv) the steady state is the state with eigenstate zero of the total thermal Hamiltonian of the system.

Having discussed  general theoretical aspects of the approach, let us consider an  important specific example of electron transport through the interacting  region. 
We begin with the general tunneling Hamiltonian:
\begin{equation}
H= H_S(a^{\dagger}_s, a_s) + \sum_{b \in L,R}  \varepsilon_b a^{\dagger}_b a_b +
\sum_{s, b \in L,R } (t_{b s} a^{\dagger}_b a_s + h.c.), 
\label{ham1}
\end{equation}
where $a^{\dagger}_b$($a_b$) creates (destroys) an electron in the state $b$ of either the left ($L$) or  the right ($R$) lead, and 
$a^{\dagger}_s$ and  $a_s$ are creation annihilation operators in the interacting region.
Here $H_S(a^{\dagger}_s, a_s) $  represents the interacting region (for example, quantum dot,  atom or molecule)  and contains two-particle electron-electron correlations and, if necessary, electron-phonon
interactions. First, we introduce the  thermal Hamiltonian by formally doubling the degrees of freedom ${\cal H} = H-\tilde{H}$.
Then, we impose the reservoir boundary conditions  by performing thermal canonical Bogoliubov transformation (\ref{bogol}) on the  particles
from the left and the right leads.
So the thermal Hamiltonian becomes:
\begin{widetext}
\begin{eqnarray}
{\cal H}=  {\cal H}_S(a^{\dagger}_s, \tilde{a}^{\dagger}_s, a_s, \tilde{a}_s)  
+ \sum_{b \in L,R} \varepsilon_b ( \beta^{\dagger}_b \beta_b - \tilde{\beta}^{\dagger}_b \tilde{\beta}_b) +
  \sum_{s, b\in L,R} t_{bs} \left[\sqrt{1-n_b} (\beta^{\dagger}_b a_s -  \tilde{\beta}^{\dagger}_b \tilde{a}_s) 
  + \sqrt{n_b}(a^{\dagger}_s \tilde{\beta}^{\dagger}_b +  \tilde{a}^{\dagger}_s \tilde{\beta}^{\dagger}_b) + \mbox{h.c.} \right] 
  \label{ham2}
\end{eqnarray}
\end{widetext}

By means of the canonical Bogoliubov transformation  for reservoir operators, 
we put all boundary conditions inside the thermal Hamiltonian.
The thermal Hamiltonian (\ref{ham2}) depends now on temperatures and chemical potentials of the leads.
 It contains terms  as $a \beta$, $a^{\dagger} \beta^{\dagger}$ etc which usually appear in the theory of superfluidity (so-called "dangerous Bogoliubov diagrams").\cite{bogoliubov58-hfb} Due to these terms  the structure of the nonequilibrium steady state for the electron transport problem will have some mathematical similarities with the vacuum in the Hartree-Fock-Bogoliubov theory.\cite{bogoliubov58-hfb}
 
Let us apply Wick theorem to the thermal Hamiltonian of the interacting region
\begin{equation}
{\cal H}_S(a^{\dagger}_s, \tilde{a}^{\dagger}_s, a_s, \tilde{a}_s) = {\cal H}^0_S + :{\cal H}_S :
\label{hs1}
\end{equation}
Here $ {\cal H}^0_S$ is the quadratic part of the  thermal Hamiltonian and $ :{\cal H}_S:$
is the normal ordered part which contains irreducible to the quadratic  two-particle interactions. The normal ordering is performed with respect to nonequilibrium vacuum which is yet to be defined.
We partition the total thermal Hamiltonian
into "simple" part ${\cal H}_0$, which collects all quadratic terms from the systems, leads and system-lead interactions, and the normal ordered part of the system thermal Hamiltonian:
$
{\cal H} = {\cal H}_0 + :{\cal H}_S:
$.
The "simple" part of the thermal Hamiltonian  ${\cal H}_0$ can be exactly diagonalized by  some  unitary rotation $ \mathbf{U}$ of annihilation and creation operators.
Then, if we combine this unitary rotation  $ \mathbf{U}$ and the thermal Bogoliubov transformation (\ref{bogol}), the total transformation from the "bare" operators to nonequilibrium quasiparticle operators can be written 
   schematically as 
 \begin{eqnarray}
\alpha =
 \mathbf{W} 
\left( 
\begin{array}{c}
a_b\\
a_s 
\end{array}
 \right)
\mathbf{W} ^{\dagger},
\label{quasiparticle}
\end{eqnarray}
 where $\alpha^{\dagger}$ ($\alpha$) are creation (annihilations) operators for nonequilibrium quasiparticles.
 The unitary transformation operator $\mathbf{W} $ takes into account both, nonequilibrium reservoir boundary conditions 
($\cos \theta_b = \sqrt{1-n_b}$) and many particle correlations:
 \begin{eqnarray}
\mathbf{W} =
 \mathbf{U} 
 \exp{\left( \begin{array}{cc}
  {-\sum_{b \in L, R} \theta_b ( a^{\dagger}_b \tilde{a}_b^{\dagger} -  \tilde{a}_b a_b)} &0 \\
  0&1 
 \end{array}
 \right) }
\label{w}.
\end{eqnarray}
The operator exponent in this transformation represents the canonical Bogoliubov transformation (\ref{bogol}) performed on the operators in the left and the right leads.
 The total thermal Hamiltonian in the nonequilibrium quasiparticle basis becomes
\begin{equation}
{\cal H} = \sum_{k} E_k (\alpha^{\dagger}_k  \alpha_k -  \tilde{\alpha}^{\dagger}_k  \tilde{\alpha}_k) + :{\cal H}_S:,
\label{ham4} 
\end{equation}
where $E_k$ is the spectrum of nonequilibrium quasiparticles.
The steady state nonequilibrium wavefunction $|0\rangle$ for  the "simple" part of the thermal 
Hamiltonian ${\cal H}_0$
(${\cal H}_0 |0\rangle =0$) is also a vacuum for nonequilibrium
quasiparticles $\alpha$ 
\begin{equation}
\alpha |0\rangle =\tilde{\alpha} |0\rangle = 0.
\label{vacuum}
\end{equation}
Now we know nonequilibrium steady state  vacuum and corresponding nonequilibrium quasiparticles, we can 
construct the nonequilibrium Fock space for our system:
\begin{equation}
|0\rangle, \;\; \alpha^{\dagger} |0\rangle, \;\; 
\tilde{\alpha}^{\dagger} |0\rangle, \;\; \alpha^{\dagger}  
\tilde{\alpha}^{\dagger}  |0\rangle, ....
\label{fock}
\end{equation}
We note here that the thermal Hamiltonian, by itself, does not mix original and "tilde" Fock spaces,
the nonequilibrium mixing is produced by so-called vacuum correlation since we introduce the common vacuum 
(\ref{vacuum}) for both spaces.
The somewhat different idea  of defining a nonequilibrium Fock space, so called third quantization, has been recently proposed by Prosen based on  quadratic Lindblad master equation.\cite{prozen08}. 

We would like to emphasize  here the following important point. Fifty years ago Bogoliubov and Landau  introduced one of the most beautiful and useful concepts in theoretical physics, a quasiparticle.\cite{landau56,bogoliubov58}
 The basic idea of quasiparticles is to represent true ground state of interacting many particle systems as  a vacuum with respect to some quasiparticle annihilation operators. 
 The quasiparticle description is, in principle, exact if one defines  the quasiparticles in terms of the exact eigenstates of many-particle systems. Although in practice it can 
be done only approximately by establishing  the relation between quasiparticles and "bare" particles of the system, for example, by  canonical Bogoliubov transformations.
Here, we demonstrated how one  can define quasiparticles  for an  interacting quantum many-particle system in nonequilibrium  in such a way that they include both 
nonequilibrium and correlations into their structure.

With the nonequilibrium Fock space and the thermal Hamiltonian, which is defined in 
this Fock space, we can readily extend  standard methods of electronic structure calculations, such 
as many-body perturbation theory, coupled-cluster technique, renormalization 
group, or configuration interaction, to  nonequilibrium steady state  systems. 
For example, the coupled cluster equations  for nonequilibrium electron transport can be easily formally obtained if we 
make the following ansatz for correlated nonequilibrium steady state 
wavefunction 
$
|\psi \rangle = \exp(S)  |0\rangle,
$
where $  |0\rangle $ is the vacuum for nonequilibrium quasiparticles and 
$
S = \sum_{ij}S_{ij} a^{\dagger}_i a_j 
+  \sum_{ijkl}S_{ijkl} a^{\dagger}_i a^{\dagger}_j a_k a_l + 
...
$ .\cite{bartlett:291}
Here the summation  is restricted to the subspace 
of the interacting region. Then the equations for coefficients $S$ can be obtained by the 
standard methods \cite{bartlett:291}  with the use of the thermal Hamiltonian (\ref{ham4}). 

In order to compare our approach with existing theoretical methods, we consider  a single impurity connected 
to two leads held at different chemical potentials. Here we can perform most calculations analytically and compare the final expressions with those obtained with nonequilibrium Green's functions. We begin with the thermal Hamiltonian (\ref{ham2}) where the system comprises only one site: 
\begin{equation}
{\cal H}_S(a^{\dagger}_s, \tilde{a}^{\dagger}_s, a_s, \tilde{a}_s) 
= \varepsilon (a^{\dagger} a -  \tilde{a}^{\dagger} \tilde{a}).
\end{equation}
 We use Heisenberg equations of motion for all system and leads creation/annihilation operators $i \dot{A} = [ A, {\cal H}]$, where ${\cal H}$ is the total thermal Hamiltonian. Then we perform the Laplace transform $f(z) = \int_0^{\infty} dt \exp(-zt) f(t)$ for all time dependent operators in these equations. 
Working with the Laplace transform it is more convenient  to assume that the system was prepared in equilibrium at $t=0$ and the nonequilibrium steady state is established at 
$t=+\infty$. In this case  we are able to compute directly the asymptotic operators by simply multiplying the Laplace transformed operator by $z$ and then letting $z$ go to zero:
$
\lim_{t\rightarrow +\infty} f(t) = \lim_{z\rightarrow 0} z f(z)
$.
We eliminate all bath operators  from the Heisenberg equations of motion and the system particle annihilation operator becomes
\begin{eqnarray}
a(z) &=& i G(z) a   
\nonumber
\\
&+&  iG(z) \sum_b \frac{t_b}{iz -\varepsilon_b} (\sqrt{1-n_b} \tilde{\beta}_b   + \sqrt{n_b} \tilde{\beta}^{\dagger}_b),
\label{az}
\end{eqnarray}
where $a$, $\tilde{\beta}$, and $\tilde{\beta}^{\dagger}$ in the right side of this equation are time-dependent operators computed at $t=0$, $a(z)$ is the Laplace transform of the system particle annihilation operator $a(t)$, and 
$G(z) =[ i z - \varepsilon - \Sigma(z)]^{-1}$ is the system Green's function. $ \Sigma(z)$, which enters the Green's function, is the standard self-energy of the leads
$\Sigma(z) = \Sigma_L(z) + \Sigma_R (z) $.
  Let us now compute, as an example, the steady state electron density on the impurity
\begin{eqnarray}
\rho =  \lim_{z\rightarrow 0} z \int^{\infty}_0 dt \exp(-z t) \langle a^{\dagger}(t) a(t) \rangle ,
\label{rho1}
\end{eqnarray}
where averaging $\langle ...\rangle$ is performed over the vacuum for  $a$, ${\beta}$, and $\tilde{\beta}$ operators. If we substitute the expression for $a(z)$ (\ref{az}) into (\ref{rho1}), we obtain
\begin{equation}
\rho =\frac{1}{2\pi i} \int_{-\infty}^{+\infty} d\omega |G(i \omega)|^2 X(\omega),
\label{rho2}
\end{equation}
where  $ X(\omega)= \sum_{B=L,R} n(\omega - \mu_B)
[ ({\Sigma}^*_B(-i \omega) - 
{\Sigma}_B(-i \omega) )]$.
The same result can be obtained by the method of nonequilibrium Green's functions
through the calculation of  G-lesser Green's function $G^{<}$  for the system (see, for example\cite{Xue02}). Likewise, the very similar calculations can be performed for the current operator (not shown here), and the
the standard Landauer formula can be recovered.

In conclusion, based on the formalism of TFD we have proposed the concept of nonequilibrium quasiparticles
and nonequilibrium Fock  space by representing a  nonequilibrium steady state of interacting many particle systems 
as  a vacuum with respect to some quasiparticle annihilation operators.  
We demonstrated that these nonequilibrium quasiparticles include correlations and nonequilibrium into their structure.
We  developed a general 
theory for the quantum many-particle systems connected to several thermal baths or particle reservoirs
as well as the general theoretical framework to deal with electron transport through the interacting regions.
We have discussed how the standard methods of electronic structure calculations such as coupled cluster theory can be extended to 
electron transport problem.
On the example of electron transport through the single level impurity we have demonstrated that the 
results obtained from our approach are consistent with those obtained from the nonequilibrium Green's functions.

The author thanks Maxim Gelin for many valuable discussions. This work has been supported by the Francqui Foundation and by the Belgian Federal Government under the
Inter-university Attraction Pole project NOSY P06/02.

\end{document}